\documentclass[a4paper,11pt]{article}
\usepackage{jinstpub} % for details on the use of the package, please see the JINST-author-manual
\usepackage[arrowdel]{physics}
\usepackage{siunitx}
\usepackage{subcaption}
\usepackage[version=4]{mhchem}
\usepackage{placeins}
\usepackage{afterpage}
\usepackage{tikz}
\usetikzlibrary{calc}
\usepackage{bm}
\usepackage{tabularray}

\usepackage{xcolor}

\DeclareSIUnit\days{days}
\DeclareSIUnit\cps{cps}
\DeclareSIUnit\elementarycharge{\text{\ensuremath{e}}}

\newcommand*\annotatedFigureBoxCustom[8]{\draw[#5,thick,rounded corners] (#1) rectangle (#2);\node at (#4) [fill=#6,thick,shape=circle,draw=#7,inner sep=2pt,font=\sffamily,text=#8] {\textbf{#3}};}
\newcommand*\annotatedFigureBox[4]{\annotatedFigureBoxCustom{#1}{#2}{#3}{#4}{white}{white}{black}{black}}

\newenvironment {annotatedFigure}[1]{\centering\begin{tikzpicture}
    \node[anchor=south west,inner sep=0] (image) at (0,0) { #1};\begin{scope}[x={(image.south east)},y={(image.north west)}]}{\end{scope}\end{tikzpicture}}
\title{Investigations of Charge Collection and Signal Timing in a multi-pixel Silicon Drift Detector}

\author[a, b, c, 1]{C.~Forstner\note[1]{Corresponding author.},}
\author[a, d]{K.~Urban,}
\author[d, e]{M.~Carminati,}
\author[a]{F.~Edzards,}
\author[d, e]{C.~Fiorini,}
\author[a]{M.~Lebert,}
\author[c]{P.~Lechner,}
\author[a]{D.~Siegmann,}
\author[a, b, c]{D.~Spreng,}
\author[a, b]{and S.~Mertens}

\affiliation[a]{Technical University of Munich, TUM School of Natural Sciences, Physics Department, James-Franck-Str.~1, 85748 Garching, Germany}
\affiliation[b]{Max Planck Institute for Nuclear Physics, Saupfercheckweg~1, 69117 Heidelberg, Germany}
\affiliation[c]{Semiconductor Laboratory of the Max Planck Society, Isarauenweg~1, 85748 Garching, Germany}
\affiliation[d]{Politecnico di Milano, Dipartimento di Elettronica, Informazione e Bioingegneria, Via C. Golgi~40, 20133 Milan, Italy}
\affiliation[e]{INFN, Sezione di Milano, Via Giovanni Celoria~16, 20133 Milan, Italy}

\emailAdd{christian.forstner@tum.de}

\abstract{Sterile neutrinos are a minimal extension of the Standard Model of particle physics and a promising candidate for dark matter if their mass is in the~\si{\keV}-range.
The Karlsruhe Tritium Neutrino experiment~(KATRIN), equipped with a novel multi-pixel silicon drift detector array, the TRISTAN detector, will be capable of searching for these~\si{\keV}-scale sterile neutrinos by investigating the kinematics of the tritium~$\upbeta$-decay.
This measurement will be performed after the completion of the neutrino mass measurement campaign.
To detect a sterile neutrino signal with a high sensitivity, a profound understanding of the detector response is required.
In this work, we report on the characterization of a 7-pixel TRISTAN prototype detector with a laser system.
We present the experimental results obtained in high-resolution scans of the detector surface with a focused laser beam and demonstrate how the charge collection and the timing of the signals generated in the detector is related to the detector geometry.
A comparison of the experimental data with simulations shows a good agreement.
}

\keywords{Detector alignment and calibration methods (lasers, sources, particle-beams); Detector modelling and simulations II (electric fields, charge transport, multiplication and induction, pulse formation, electron emission, etc); Interaction of radiation with matter; Solid state detectors}

\arxivnumber{2409.08901}

\begin{document}
\maketitle
\flushbottom

\section{Introduction}
\label{sec:intro}

Sterile neutrinos\footnote{The term \textit{sterile neutrino} refers to an additional fourth neutrino mass eigenstate that is not purely sterile but an admixture with the three active flavor eigenstates.} are a natural and minimal extension to the Standard Model~(SM) of particle physics~\cite{Adhikari2017}.
These neutrinos could resolve various open questions in neutrino physics and cosmology depending on their production mechanism~\cite{Abazajian2012, Domcke2021}.
For a sterile neutrino mass~$m_{4}$ in the~\si{\keV}-range, these particles would be a promising dark matter candidate~\cite{Boyarsky2019}.
The mixing of active and sterile neutrinos is characterized by the active-sterile mixing amplitude~$\sin^{2}{\Theta}$, which allows to search them in the single $\upbeta$-decay of unstable isotopes like tritium.

The Karlsruhe Tritium Neutrino experiment~(KATRIN) is designed to measure the effective electron antineutrino mass using the kinematics of the tritium~$\upbeta$-decay with a final sensitivity of better than~\SI{0.3}{\eV / c^{2}}~(\SI{90}{\%}~C.L.) after a total measurement time of~\SI{1000}{\days}~\cite{Aker2021}.
After the completion of the neutrino mass measurements in~2025, the experimental apparatus will be modified to enable the search for sterile neutrinos in the~\si{\keV}-mass range with a targeted sensitivity at the parts-per-million~(ppm) level~(i.\,e.\ $\sin^{2}{\Theta} < 10^{-6}$~\cite{Mertens2019}) including systematic uncertainties to improve existing laboratory limits by several orders of magnitude~\cite{Hiddemann1995, Holzschuh1999, Kraus2013, Belesev2013, Abdurashitov2017}.
While the imprint of the neutrino mass is a distortion near the kinematic endpoint of the tritium spectrum at~$E_{0} = \SI{18.6}{\keV}$, the sterile neutrino would manifest itself as a spectral distortion with a kink-like signature at an energy~$E_{0} - m_{4} \si{c^{2}}$.
As this feature could be located anywhere in the energy spectrum, the measurement range is extended from the endpoint region to the entire spectrum.
This extension is achieved by lowering the retarding potential of the~KATRIN main spectrometer which utilizes the magnetic adiabatic collimation with electrostatic~(MAC-E) filter principle~\cite{Lobashev1985, Picard1992, Beamson1980}. 
Only electrons with sufficient energy to overcome this potential are capable of reaching the detector.
By reducing the voltage applied to the main spectrometer, the number of transmitted electrons will increase by up to 12 orders of magnitude~\cite{Mertens2015}.
While in the current~KATRIN measurement mode an integral tritium spectrum is obtained by measuring the count rate of the $\upbeta$-electrons as a function of the retarding potential, a differential measurement of the $\upbeta$-spectrum is planned in which the energy of each electron is determined by the detector~\cite{Mertens2019}. 
Since the current focal-plane detector system~\cite{Amsbaugh2015} of~KATRIN is not designed to handle the high count rates involved~\cite{KATRIN2005, Aker2021}, a novel detector system is required.
In an optimized setting~\cite{Mertens2019}, where the~KATRIN tritium source is operated at~\SI{1}{\%} of its nominal density, the detector needs to be capable of handling the high count rates of up to~\SI[print-zero-exponent=true, print-unity-mantissa=false, exponent-mode=scientific]{1e{8}}{counts\ per\ second~(\cps)} of the strong tritium source.
In addition, to resolve the spectral distortion the detector system needs an excellent energy resolution of~\SI{300}{\eV} full width at half maximum~(FWHM) at~\SI{20}{\keV}.
These requirements are fulfilled by the TRISTAN detector and readout system currently being developed~\cite{Mertens2020, Siegmann2024}.
The TRISTAN detector is based on the Silicon Drift Detector~(SDD) technology~\cite{Gatti1984a, Gatti1984b, Rehak1985}, which features a very low anode capacitance and therefore allows for high-precision electron spectroscopy at high rates~\cite{Lechner2001}.

For the sterile neutrino search, a comprehensive understanding of the detector response to the incident radiation is essential.
Sensitivity studies are conducted to model the impact of a variety of systematic effects on the obtained $\upbeta$-spectrum originating from the different components of KATRIN~(rear wall, tritium source, transport and spectrometer section, detector, and data acquisition system).
Since about~\SI{1}{\%} of all detected events are charge sharing events, they have to be accounted for in the model describing the detector response.
Charge sharing refers to the process in which the generated charge carriers are collected in multiple pixels.
It is the result of energy being deposited at or close to the boundary region of adjacent detector pixels.
In this work, we focus on investigating this effect, which can be parameterized by the effective charge cloud size~$w_{\mathrm{cc}}$.
Moreover, we study the drift time of the electrons as a function of the position at which they are generated.
This allows us to assess the timing resolution of the detector and to find a possible correlation of the drift time to events which share their charge between adjacent pixels.
Furthermore, we examine the pulse shape to infer the location of energy deposition and thus to identify charge sharing events.
The results will be used as input parameters for the sensitivity studies and will be incorporated into the model describing the detector response.

For the investigations, a laser system has been designed to characterize the charge collection process in a 7-pixel~TRISTAN prototype~SDD, see sections~\ref{sec:detector} and~\ref{sec:setup}.
It allows to mimic electron signals by creating electron-hole pairs at a precise position in the pixel at a shallow depth, where also electrons would deposit their energy.
The experimental results will be compared to simulations, which are introduced in section~\ref{sec:simulations}.
An empirical model describing the pulse shape has been developed to compare the charge collection process at different positions in the detector, see section~\ref{sec:signal_model}.
Finally, the experimental results will be presented in section~\ref{sec:results}.

We note that the timing performance of~SDDs has been previously investigated using a similar experimental approach employing~SDDs with \SI{2}{\milli\meter} and \SI{5}{\milli\meter}-side square pixels~\cite{Giacomo2022}.
However, these~SDDs not only differ in geometry, but also use a different electronic readout circuit than the one used in this work.
Furthermore in~\cite{Giacomo2022}, the focus was put on obtaining the timing resolution of~SDDs as the uncertainty in the measurement of the zero-crossing time of a bipolar output pulse.
In contrast, the objective of the present study is to examine the drift time of electron charge clouds and to investigate the variation of the signal rise time as a function of different positions irradiated with the laser.

\section{TRISTAN detector and readout system}
\label{sec:detector}

\begin{figure}[t]
    \centering
    \begin{subfigure}{0.49\textwidth}
        \centering
        \includegraphics[width=0.68\textwidth,keepaspectratio,angle=90]{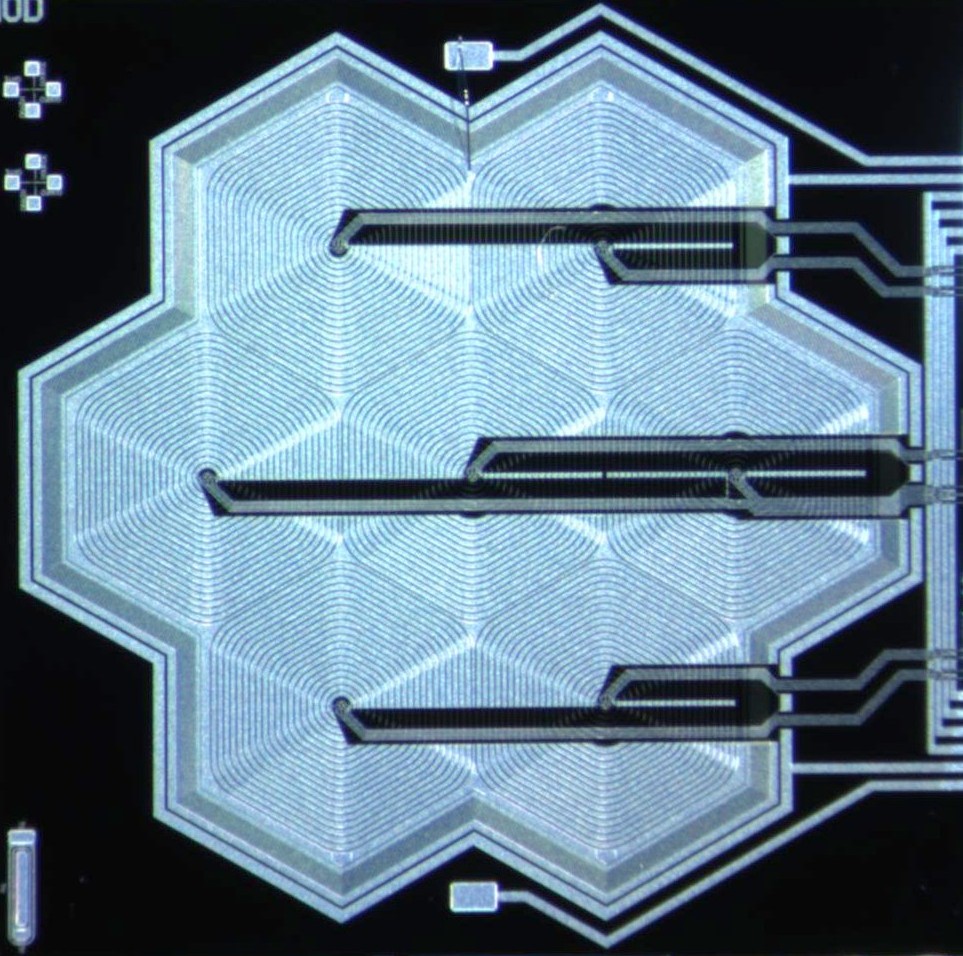}
        \caption[Readout side of a 7-pixel TRISTAN SDD]{\textbf{Readout side of a 7-pixel TRISTAN SDD}}
        \label{fig:tristan_seven_pixel_detector}
    \end{subfigure}
    \hfill
    \begin{subfigure}{0.49\textwidth}
        \centering
        \includegraphics[height=0.17\textheight,keepaspectratio,angle=90]{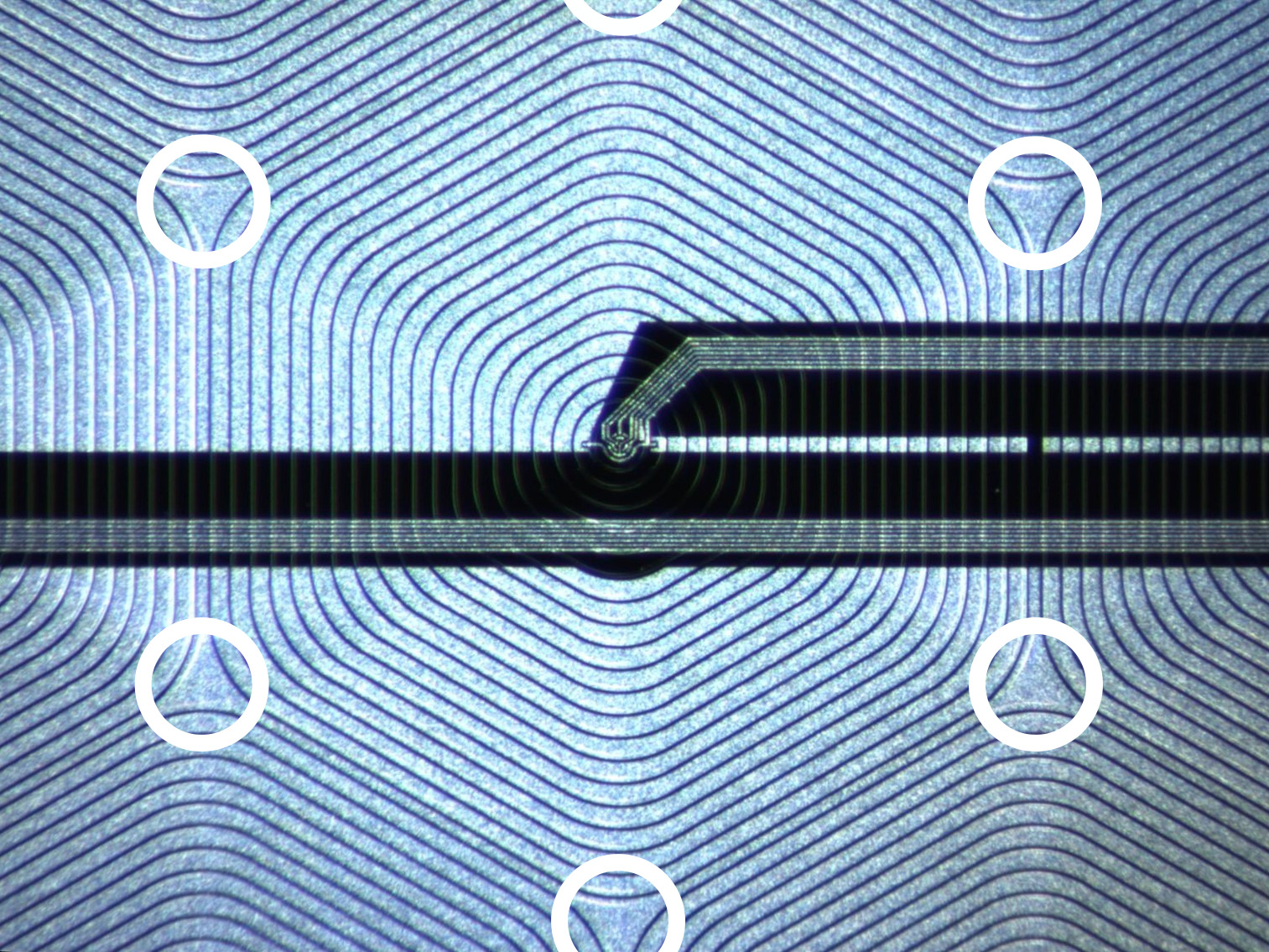}
        \caption[Close up of the central detector pixel]{\textbf{Close up of the central detector pixel}}
        \label{fig:tristan_detector_close_up}
    \end{subfigure}
    \caption[Microscope images of a 7-pixel TRISTAN SDD]{\textbf{Microscope images of a 7-pixel TRISTAN SDD}\\
    (a): Readout side of a 7-pixel TRISTAN SDD.
    The hexagonal arrangement of the drift rings forms the individual detector pixels.
    The vertical metal bus structure carries the signal and power lines.
    The size of the chip is $\SI{11}{\milli\meter} \times \SI{12}{\milli\meter}$.\\
    (b): Close up of the central detector pixel.
    The 20 hexagonal drift rings and, in the center, the electrode structure required for pixel biasing and readout can be noticed.
    The outermost drift ring is shared between all outer pixels due to the monolithic detector design.
    The triangular patches (highlighted by white circles) define the triple points where three adjacent pixels are in contact.
    }
    \label{fig:tristan_detector}
\end{figure}

The TRISTAN detector makes use of the Silicon Drift Detector~(SDD) technology originally designed for X-ray spectroscopy~\cite{Lechner1996}.
The detectors are multi-pixel SDD arrays of~\SI{450}{\micro\meter}~thickness and are fabricated from a monolithic n-type silicon wafer with a high resistivity of about~\SI{4}{\kilo\ohm \centi\meter} by the Semiconductor Laboratory~(HLL) of the Max Planck Society.
The hexagonal pixels consist of 20 drift rings each and are arranged in a honeycomb pattern to have no insensitive detection area in between.
Each pixel has a circumscribed diameter of~\SI{3.3}{\milli\meter} and a surface area of about~\SI{7}{\milli\meter^{2}}.
Figure~\ref{fig:tristan_detector} shows a photograph of a 7-pixel TRISTAN SDD.
In order to take advantage of the low anode capacitance of~$\mathcal{O}(\SI{100}{\femto\farad})$~\cite{Lechner1996}, an n-channel junction-gate field-effect transistor~(JFET) is integrated into the detector chip near the anode of each pixel~\cite{Lechner2001}.
This~JFET is the central part of the charge-sensitive pre-amplifier~(CSA), which is completed by an external application-specific integrated circuit~(ASIC), named~ETTORE~\cite{Trigilio2018}.
The concept of an internal~JFET provides an adequate noise performance and allows the electronics to be installed in some distance to the anodes which is advantageous for larger pixel arrays.

An event occurs in the detector when radiation enters the detector volume on the entrance window side~(opposite to the readout side) and excites the silicon atoms of the material.
The deposited energy is used to create electron-hole pairs which form two oppositely charged clouds.
The charge carriers are subject to the electric field in the~SDD and drift to the respective electrodes - the electrons to the anode and the holes to the cathodes.
Since in the TRISTAN detector only the anode of a detector pixel is read out, we only focus on the electrons in the following investigations.

\FloatBarrier

\section{Experimental setup}
\label{sec:setup}

\begin{figure}[t!]
    \centering
    \begin{subfigure}{0.49\textwidth}
        \centering
        \begin{annotatedFigure}
            {\includegraphics[width=\textwidth,keepaspectratio]{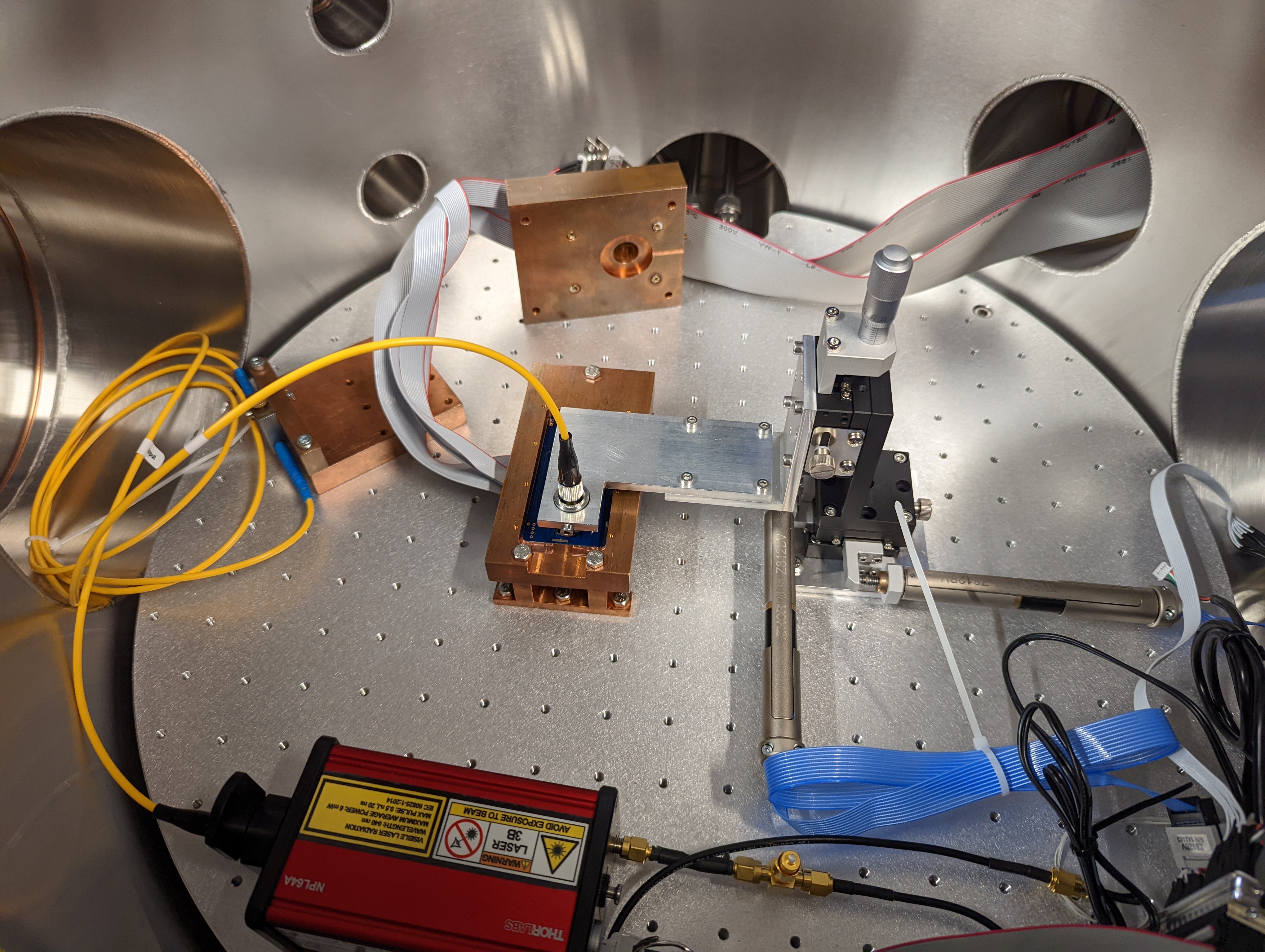}}
            \annotatedFigureBox{0.15, 0.01}{0.52, 0.25}{\small 1}{0.52, 0.25}
            \annotatedFigureBox{0.01, 0.35}{0.27, 0.67}{\small 2}{0.27, 0.35}
            \annotatedFigureBox{0.35, 0.36}{0.55, 0.63}{\small 3}{0.35, 0.63}
            \annotatedFigureBox{0.59, 0.2}{0.8, 0.8}{\small 4}{0.59, 0.8}
        \end{annotatedFigure}
        \caption[Overview of the entire setup]{\textbf{Overview of the entire setup}}
        \label{fig:detector_setup_overview}
    \end{subfigure}
    \hfill
    \begin{subfigure}{0.49\textwidth}
        \centering
        \begin{tikzpicture}
            \node[anchor=south west,inner sep=0] (image) at (0,0) {
                \includegraphics[width=\textwidth,keepaspectratio]{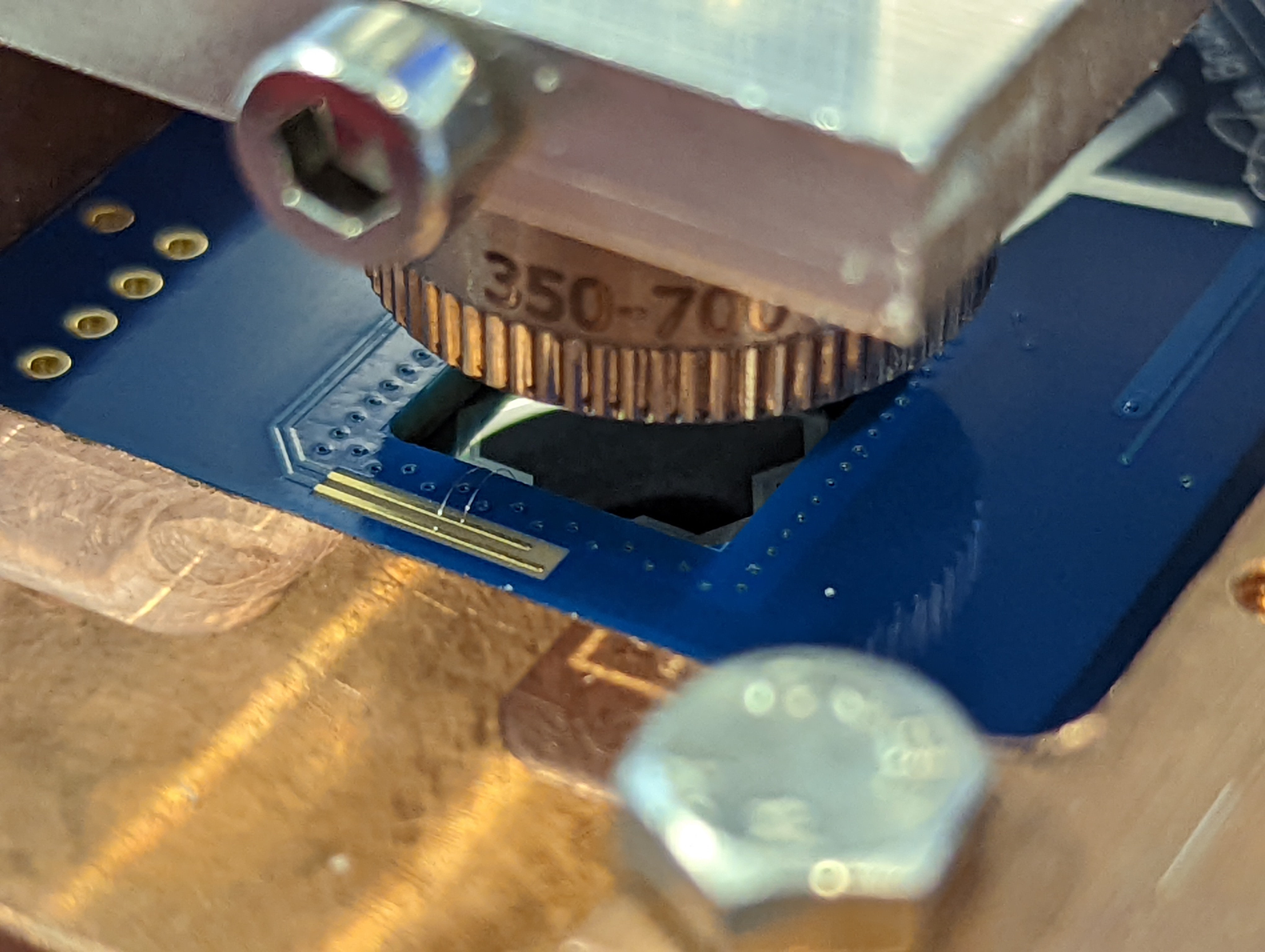}
            };
            \begin{scope}[
                x={($0.1*(image.south east)$)},
                y={($0.1*(image.north west)$)}]
                    \draw[rounded corners,latex-,very thick,white,font=\small] (8.0,5.0) -- ++(1.0,2.0) node[above,black,fill=white]{PCB};    
                    \draw[rounded corners,latex-,very thick,white,font=\small] (4.6,5.2) -- ++(-2.4,-2.2) node[below,text width=1.5cm,align=center,black,fill=white]{Entrance\\Window};    
                    \draw[rounded corners,latex-,very thick,white,font=\small] (5,6.5) -- ++(2.7,-4.0) node[below,black,fill=white]{Collimator};    
            \end{scope}
        \end{tikzpicture}
        \caption[Close up of the detector entrance window]{\textbf{Close up of the detector entrance window}}
        \label{fig:detector_setup_close_up}
    \end{subfigure}
    \caption[Experimental setup for the detector characterization]{\textbf{Experimental setup for the detector characterization}\\
    (a): The laser~(1) emits a Gaussian beam which is transmitted via the yellow optical fiber with an integrated attenuator~(2) and collimated onto the detector entrance window by the lens mounted above in an L-shaped optical table~(3).
    This table is attached to an XYZ linear translation stage~(4) whose movements are automatized using two actuators which allows for high-precision scans of the SDD.\\
    (b): The detector printed circuit board (PCB) is mounted facing upwards with the collimator positioned directly above the entrance window of the SDD.
    The laser beam is focused on the SDD and moved over the detector pixels.}
    \label{fig:detector_setup}
\end{figure}

For the characterization of the charge collection and the signal timing of the 7-pixel TRISTAN SDD, a dedicated experimental setup has been developed.
It consists of a laser, whose beam is directed onto the entrance window of the detector using fiber optics.
The system is placed in a light-tight vacuum chamber at room temperature~($T = \SI{300}{\kelvin}$) and is shown in figure~\ref{fig:detector_setup}.

A pulsed laser~(Thorlabs NPL64A) with a wavelength of~$\lambda = \SI{640 \pm 10}{\nano\meter}$ emits photons with an energy of about~\SI{1.9}{\eV}.
The intensity profile of the laser beam is considered to represent an ideal Gaussian beam with standard deviation~$w_{\mathrm{laser}} = \SI{6.3 \pm 0.7}{\micro\meter}$~\cite{Kogelnik1965}.
This matches our requirements of $w_{\mathrm{laser}} < \SI{15}{\micro\meter}$ to determine the effective size of the electron charge cloud, see section~\ref{subsec:cc_width}.
At the given energy of~\SI{1.9}{\eV}, photons penetrate silicon up to~\SI{10}{\micro\meter}~\cite{Green1995}.
This is similar to the range of electrons emitted in the decay of tritium during the~\si{\keV}-scale sterile neutrino search, as their kinetic energy is increased by a post-acceleration electrode~(PAE) by up to~\SI{20}{\keV} before impinging on the detector surface.
Consequently, the laser used for the investigations in the scope of this work is well suited to mimic an electron signal in the detector.

Using an external pulse generator, the laser is triggered with a frequency of~\SI{500}{\kilo\hertz} and provides rectangular pulses with a typical width of~\SI{10 \pm 1}{\nano\second}.
These settings are chosen to allow both, a high laser repetition rate for a high statistics measurement, and a sufficiently long time interval between each pulse for the detector to return to its nominal, fully depleted state within a reasonable measurement time.

The laser beam is transmitted via a single-mode optical fiber with an integrated attenuator~(Thorlabs VOA630-FC) which allows to manually vary the attenuation of the beam in the fiber.
The attenuation is chosen such that only a fraction of the emitted photons is transmitted to the detector.
By calibrating the detector with an \ce{^{55}Fe} source with its two prominent X-ray lines at~\SI{5.90}{\keV}~(Mn-K$_{\alpha}$) and~\SI{6.49}{\keV}~(Mn-K$_{\beta}$)~\cite{Junde2008}, the deposited energy was tuned to be~$E = \SI{10.6}{\keV}$ which corresponds to the transmission of around~\SI{9600} photons per pulse.
The bias voltages applied to operate the detector are listed in table~\ref{tab:bias_voltages}.
\begin{table}[t]
    \centering
    \caption[External bias voltages for the operation of the detector]{\textbf{External bias voltages for the operation of the detector}\\
    The relevant bias voltages applied to the electrodes on the detector readout side are listed.
    The voltages were optimized to achieve the best possible energy resolution of \SI{181.2 \pm 1.7}{\eV} across all detector pixels at the \SI{5.90}{\keV}~(Mn-K$_{\alpha}$) X-ray line at room temperature.}
    \begin{tblr}{c|c|c|c|c|c|c}
        \hline
        Electrode & $V_{\mathrm{D}}$ & $V_{\mathrm{IGR}}$ & $V_{\mathrm{R1}}$ & $V_{\mathrm{RX}}$ & $V_{\mathrm{BC}}$ &$V_{\mathrm{BF}}$\\
        \hline
        Voltage & \SI{6.6}{\volt} & \SI{-19.0}{\volt} & \SI{-9.0}{\volt} &  \SI{-120.0}{\volt} & \SI{-100.0}{\volt} & \SI{-110.0}{\volt}\\
        \hline
    \end{tblr}
    \label{tab:bias_voltages}
\end{table}
As the laser is not thermally regulated during operation, fluctuations of the laser output power have been observed over time leading to variations in the deposited energy of less than~\SI{5}{\%}.

Subsequently, the beam is focused using a collimator~(Thorlabs CFC2-A) which is equipped with a lens of~\SI{2}{\milli\meter} focal length.
The collimating lens is mounted in an L-shaped optical table facing downwards into the direction of the detector entrance window.
This optical table is attached to an \textit{XYZ} linear translation stage, that is controlled via linear motorized actuators~(Thorlabs Z812BV).

By adjusting the vertical distance between the lens and the detector, the laser beam is directed onto the entrance window of the~SDD.
The optimal distance has been determined by focusing the beam onto the sensor of a USB~webcam, whereby the minimum beam width~$w_{\mathrm{laser}}$ has been obtained.
After the optimization in the \textit{Z} direction, the corresponding stage has been fixed.
In contrast, the \textit{X} and \textit{Y} translation stages are kept freely movable.

A two-dimensional grid of \textit{X} and \textit{Y} coordinates is used to cover the surface area of the detector.
In this work, two different detector areas have been scanned using different measurement grids, see figure~\ref{fig:detector_grid}:
\begin{figure}[t]
    \centering
    \begin{subfigure}{0.49\textwidth}
        \centering
        \includegraphics[width=\textwidth,keepaspectratio]{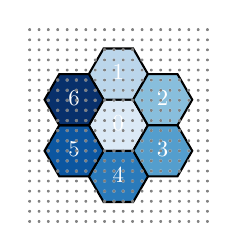}
        \caption[Exemplary grid over the entire detector]{\textbf{Exemplary grid over the entire detector}}
        \label{fig:detector_total_grid}
    \end{subfigure}
    \hfill
    \begin{subfigure}{0.49\textwidth}
        \centering
        \includegraphics[width=\textwidth,keepaspectratio]{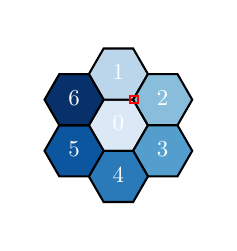}
        \caption[Triple point region]{\textbf{Triple point region}}
        \label{fig:triple_point_grid}
    \end{subfigure}
    \caption[Illustration of the grid for scanning the detector entrance window with a laser]{\textbf{Illustration of the grid for scanning the detector entrance window with a laser}\\
    (a): For the investigation of the homogeneity and the behavior of the individual detector pixels with respect to the charge collection and signal timing, the entire detector surface is scanned using a rectangular grid of~$350 \times 350$ points with a spacing of~\SI{25}{\micro\meter}.
    Here, a grid of~$20~\times~20$~points with a spacing of~\SI{0.5}{\milli\meter} is used for illustration.\\
    (b): To characterize the influence of charge sharing, a second scan with higher resolution at the region, where three detector pixels are in contact, is conducted using a grid of~$90~\times~90$~points with a spacing of~\SI{4.5}{\micro\meter}.
    This region is called a triple point and is marked by the red rectangle.
    Since the grid points would fill the entire region, the grid is not shown.
    }
    \label{fig:detector_grid}
\end{figure}
\begin{itemize}
    \item For the first measurement, a grid is generated covering the entire surface of the detector entrance window, see figure~\ref{fig:detector_total_grid}.
    The goal is to investigate the homogeneity and the behavior of the individual pixels with respect to charge collection and signal timing.
    The grid is chosen to be a square consisting of~$350~\times~350$~points with a spacing of~\SI{25}{\micro\meter}.
    This configuration provides a good compromise between the spatial resolution of scanning the detector surface and the total measurement duration of about \SI{200}{\hour}.
    \newpage
    \item In the second measurement, the area in which three detector pixels are adjacent to each other, a so-called triple point, is studied to investigate the effect of charge sharing between three neighboring pixels.
    For charge sharing to occur, charge clouds can be created in a single detector pixel, with some electrons of the clouds moving into a neighboring pixel.
    Alternatively, the clouds can already extend across the boundaries of neighboring pixels as they are generated.
    The movement of the charges is determined by the field configuration in this region and is described in section~\ref{sec:simulations}.
    The region where pixels 0, 1, and 2 are in contact is chosen to investigate this effect, see figure~\ref{fig:triple_point_grid}.
    A grid of~$90~\times~90$~points with a spacing of~\SI{4.5}{\micro\meter} is generated for a high-resolution scan lasting about~\SI{11}{\hour}.
\end{itemize}
The actuators steering the motorized stage moves the collimator to the specified position of each \textit{X-Y} coordinate.
For each grid position, \SI{8.4}{\milli\second} of waveform signal including about~$4000$ signal pulses created by the laser pulses as well as the external laser trigger pulses are recorded.
Despite the fluctuating laser output power throughout the measurements, the deposited energy of the signals created by the laser pulses recorded at a single grid coordinate has been found to be Poisson distributed as it is expected.
By detecting the rising edge of the individual trigger pulses, the obtained signal pulses are averaged to minimize noise fluctuations in the waveform signal.

A major advantage of using a laser to characterize the detector compared to a radioactive source is that the~SDD is irradiated at the exact location defined in the grid.
Moreover, since the laser pulses carry a certain amount of energy, a fixed number of electron-hole pairs in the silicon bulk of the detector is created.
However, the main advantage of using an external triggered laser is that it provides the precise timing information of when the photons are emitted and when they enter the detector, allowing the detailed study of the signal timing and the detector time response.

\FloatBarrier

\section{Simulations}
\label{sec:simulations}

\begin{figure}[t]
    \centering
    \captionsetup[subfigure]{justification=centering}
    \begin{subfigure}{0.9\textwidth}
        \centering
        \includegraphics[width=\textwidth,keepaspectratio]{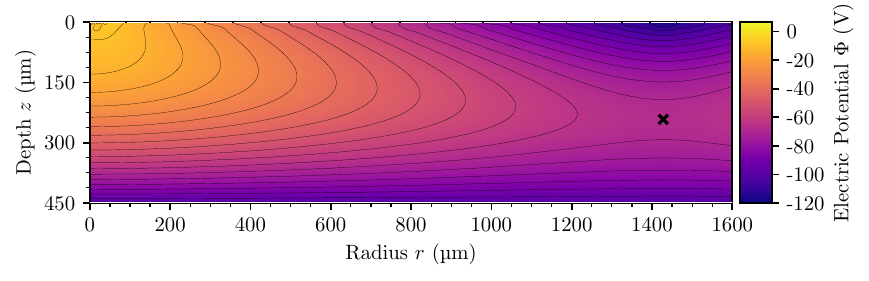}
        \caption[Simulation of electric potential]{\textbf{Simulation of the electric potential}}
        \label{fig:simulation_electric_potential}
    \end{subfigure}
    \begin{subfigure}{0.9\textwidth}
        \centering
        \includegraphics[width=\textwidth,keepaspectratio]{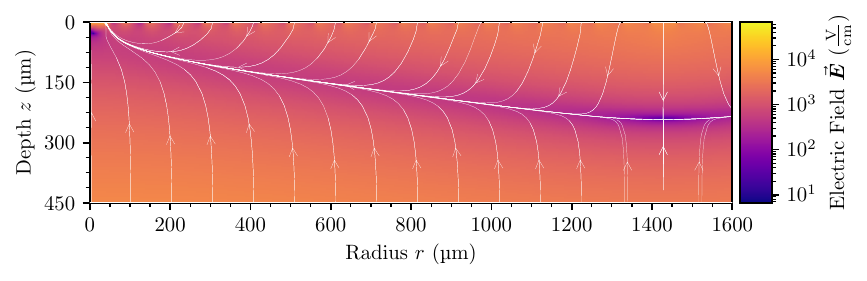}
        \caption[Simulation of electric field]{\textbf{Simulation of the electric field}}
        \label{fig:simulation_electric_field}
    \end{subfigure}
    \caption[Simulation of the field distribution in a TRISTAN SDD pixel]{\textbf{Simulation of the field distribution in a TRISTAN SDD pixel}\\
    (a): The electric potential~$\Phi$ is obtained by solving Poisson's equation as a function of the radius~$r$ and depth~$z$ of the detector pixel.
    The equipotential lines have a spacing of \SI{5}{\volt} and indicate the formation of the potential valley.
    The graph extends across the pixel border to show the saddle point~($\times$) at the pixel boundary. 
    The pixel electrodes placed on the entrance window side ($z = \SI{450}{\micro\meter}$) and on the readout side ($z = \SI{0}{\micro\meter}$) are shown in white.\\
    (b): The electric field~$\va*{E}$ determines the drift of the electrons in the silicon bulk.
    The field lines demonstrate the concept of sidewards depletion and illustrate the typical drift paths of the electrons up to their collection at the anode.
    The magnitude of the electric field is shown in color.
    }
    \label{fig:simulation_field_distribution}
\end{figure}

Dedicated detector simulations complement the experimental investigations.
The goal of the simulations is to model the transport of the electron-hole pairs generated by the incident radiation such as photons from a laser pulse in a single pixel of the TRISTAN detector and the signal formation at its readout anode.
The implementation of the simulation software consists of two separate modules:~the computation of the electric potential in a TRISTAN~SDD~pixel and the simulation of the charge carrier transport in this field.
This approach is similar to pulse shape simulations performed for Germanium detectors~\cite{Abt2021}.
Since the seven pixels of the TRISTAN detector have the same field configuration, the results are applicable to all pixels.
To obtain the electric~potential~$\Phi$ in a detector pixel, the three-dimensional Poisson's equation is solved:
\begin{equation}
    \centering
    \Delta \Phi = - \frac{N \si{\elementarycharge}}{\epsilon_{0} \epsilon_{\mathrm{Si}}} \,.
    \label{eq:poisson_equation}
\end{equation}
Here, $N$~denotes the effective doping concentration of the detector bulk material of \SI{0.9E12}{\cm^{-3}}, \si{\elementarycharge}~the elementary charge, $\epsilon_{0}$~the vacuum permittivity, and $\epsilon_{\mathrm{Si}} = \SI{11.68}{}$ the relative permittivity of silicon~(Si) at room temperature~$T = \SI{300}{\kelvin}$~\cite{Afsar1983}.
Despite their hexagonal shape, the detector pixels are considered to be radially symmetric in first order approximation, where $r$ is the radial and $z$ is the axial component of the electric potential.
Equation~\ref{eq:poisson_equation} is solved numerically using the software~\textsc{SfePy}, which implements the finite element method~\cite{Cimrman2019}.
The solution of this partial differential equation is constrained by the pixel geometry and the bias voltages applied to the individual electrodes.
Figure~\ref{fig:simulation_electric_potential} shows the resulting electric potential as a function of the pixel radius $r$ and depth $z$.
While the back contact~($V_{\mathrm{BC}}$) and back frame~($V_{\mathrm{BF}}$) electrodes are located at~$z = \SI{450}{\micro\meter}$ on the entrance window side, the integrated JFET~(including $V_{\mathrm{D}}$), an inner guard ring~($V_{\mathrm{IGR}}$), the anode, and the drift rings~($V_{\mathrm{R1}}$ to $V_{\mathrm{RX}}$) are positioned on the readout side at~$z = \SI{0}{\micro\meter}$.
More detailed information on the structure of a SDD can be found in~\cite{Lechner1996, Fiorini1999}.

The concept of sidewards depletion is illustrated by the equipotential lines forming a potential valley which guides all electrons to the pixel anode near the center at~$r \approx \SI{30}{\micro\meter}$.
Since the outermost drift ring is shared with the neighboring pixels, a saddle point is formed in the electric potential at the boundary between these pixels at~$r = \SI{1428}{\micro\meter}$.
Charge clouds created in this potential regime can remain there for a period of~$\mathcal{O}(\SI{100}{\nano\second})$ since the radial component of the electric field is minimal at the pixel borders.
Therefore, the charge carrier transport in this region is dominated by diffusion and self-repulsion processes.

The electric field is obtained as the gradient of the electric potential, $\va*{E} = -\grad{\Phi}$, and is illustrated in figure~\ref{fig:simulation_electric_field}.
Charge carriers generated by incident radiation move parallel to the direction of the electric field with an average drift velocity $\va*{v} = \mu \va*{E}$, which depends on the mobility~$\mu$ of the charge carriers.
The simulation uses a field-dependent mobility for electrons~$\mu_{\mathrm{n}}$ and holes~$\mu_{\mathrm{p}}$ in silicon at $T = \SI{300}{\kelvin}$~\cite{Jacoboni1977}.
During the drift, the charge clouds undergo diffusion processes driven by an inhomogeneous distribution of the charge carriers, resulting in a carrier concentration gradient.
As a time-dependent process, this leads to a Gaussian broadening of the charge clouds with width~$\sigma = \sqrt{2 D t}$~\cite{Fick1855}.
Here, $t$ is the time the charges drift in the detector volume and $D$ is the diffusion constant, which is related to the mobility through the Einstein–Smoluchowski equation~$D = \mu k_{\mathrm{B}} T / \si{\elementarycharge}$ with the Boltzmann constant~$k_{\mathrm{B}}$.
The self-repulsion of the charge carriers is taken into account by calculating the electric force of the individual charges acting on each other.
According to Coulomb's law, this leads to an expansion of the charge cloud, as particles with the same electrical charge repel each other.
For the simulation of the carrier transport, the Monte Carlo~(MC) method is employed.
We make use of graphics processing unit~(GPU) accelerated computing using the deep learning library~\textsc{PyTorch} to improve the simulation performance~\cite{Paszke2019}.
Given an incident energy~$E_{0}$, a fixed, integer number of electron-hole pairs~$N = E_{0} / \eta_{0}$ is generated, where~$\eta_{0}$ is the pair creation energy of~\SI{3.62}{\eV}~\cite{Knoll2010}.
In combination with the interaction position at radius~$r_{0}$, the charge carrier transport is simulated in time steps of~$\Delta t = \SI{10}{\pico\second}$.
The total simulation time is set to~\SI{1}{\micro\second}, which ensures that all charges are collected by the corresponding electrodes.
The instantaneous current induced on the anode by the electron motion is calculated using Ramo's theorem and the resulting signal pulse is obtained.
A Gaussian filter with a width of~\SI{4}{\nano\second} as well as a first-order low-pass Bessel filter with a \SIrange{10}{90}{\%}~rise time of~\SI{25}{\nano\second} are applied to mimic the response of the laser pulse and of the CSA~readout circuit.
Finally, the processed signal pulse can be compared with the experimental data.

\section{Modeling the signal shape}
\label{sec:signal_model}

\begin{figure}[t]
    \centering
    \captionsetup[subfigure]{width=\linewidth, format=hang}
    \begin{subfigure}{0.32\textwidth}
        \centering
        \includegraphics[width=\textwidth,keepaspectratio]{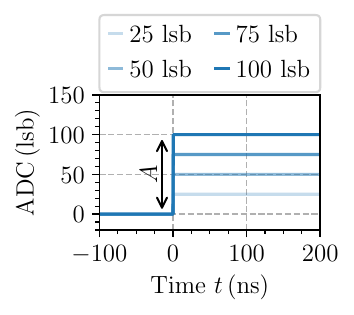}
        \caption[Model parameter: Amplitude~$\bm{A}$]{\textbf{Model parameter:\\Amplitude~$\bm{A}$}}
        \label{fig:model_amplitude}
    \end{subfigure}
    \hfill
    \begin{subfigure}{0.32\textwidth}
        \centering
        \includegraphics[width=\textwidth,keepaspectratio]{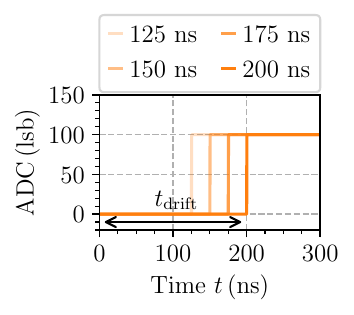}
        \caption[Model parameter: Drift time~$\bm{t{\mathrm{drift}}}$]{\textbf{Model parameter:\\Drift time~$\bm{t_{\mathrm{drift}}}$}}
        \label{fig:model_drift_time}
    \end{subfigure}
    \hfill
    \begin{subfigure}{0.32\textwidth}
        \centering
        \includegraphics[width=\textwidth,keepaspectratio]{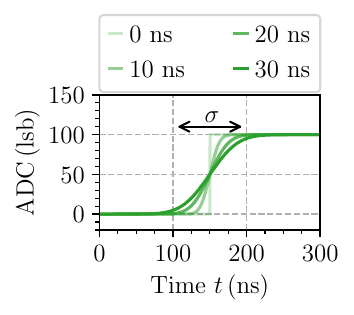}
        \caption[Model parameter: Gaussian width~$\bm{\sigma}$]{\textbf{Model parameter:\\Gaussian width~$\bm{\sigma}$}}
        \label{fig:model_gaussian_width}
    \end{subfigure}
    \hfill
    \begin{subfigure}{0.32\textwidth}
        \centering
        \includegraphics[width=\textwidth,keepaspectratio]{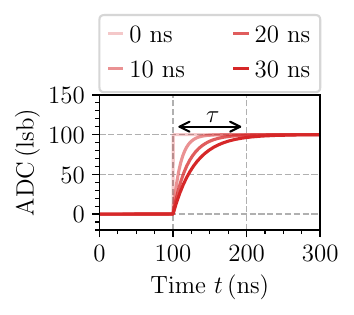}
        \caption[Model parameter: Exponential rise time~$\bm{\tau}$]{\textbf{Model parameter:\\Exponential rise time~$\bm{\tau}$}}
        \label{fig:model_exp_rise_time}
    \end{subfigure}
    \hfill
    \begin{subfigure}{0.32\textwidth}
        \centering
        \includegraphics[width=\textwidth,keepaspectratio]{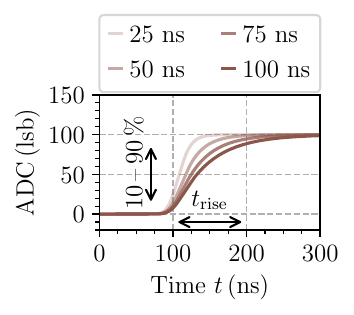}
        \caption[\SIrange{10}{90}{\%}~signal rise time~$\bm{t_{\mathrm{rise}}}$]{\textbf{\SIrange{10}{90}{\%}~signal rise time~$\bm{t_{\mathrm{rise}}}$}}
        \label{fig:model_rise_time}
    \end{subfigure}
    \hfill
    \begin{subfigure}{0.32\textwidth}
        \centering
        \includegraphics[width=\textwidth,keepaspectratio]{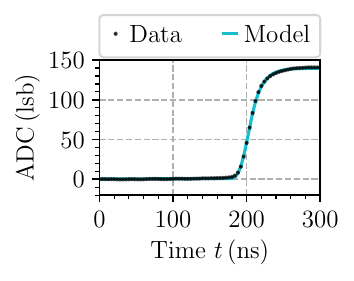}
        \caption[Example of the best fit to an averaged waveform signal]{\textbf{Example of the best fit to an averaged waveform signal}}
        \label{fig:model_fit}
    \end{subfigure}
    \caption[Impact of the various parameters on the model describing the signal shape]{\textbf{Impact of the various parameters on the model describing the signal shape}\\
    (a): The pulse height is given by the amplitude~$A$ and represents the collected charge.\\
    (b): The drift time~$t_{\mathrm{drift}}$ accounts for the spatial extension of the detector.\\
    (c): The Gaussian width~$\sigma$ describes the influence of diffusion during the drift of the charge cloud.\\
    (d): The shape of the exponential rise of the signal pulses is described by the exponential rise time~$\tau$.\\
    (e): The \SIrange{10}{90}{\%}~signal rise time~$t_{\mathrm{rise}}$ is reconstructed independently of the other model parameters, so that the results can be compared with other measurements. Here, $\sigma$ is kept constant at \SI{10}{\nano\second} and $\tau$ is varied between \SI{10}{\nano\second} and \SI{45}{\nano\second}.\\
    (f): The best fit of the model (cyan) to exemplary measurement data (black) is illustrated.
    }
    \label{fig:model_parameters}
\end{figure}

In order to compare the signal pulses of the different detector regions, an empirical model has been developed to describe the shape of the signals.
The model comprises four different parameters, each describing a physical property:
\begin{itemize}
    \item \textbf{Amplitude~\bm{$A$}}:~The amplitude describes the height of the signal pulse and therefore the amount of collected charge at the respective pixel anode.
    The amplitude is directly proportional to the energy deposited by the incident radiation.
    It is illustrated in figure~\ref{fig:model_amplitude}.
    \item \textbf{Drift~time~\bm{$t_{\mathrm{drift}}$}}:~This parameter takes into account the spatial extension of the detector pixel.
    Charge clouds have to drift through the detector bulk before they reach the anode in the center of each detector pixel.
    This is described in detail in section~\ref{subsec:drift_time} and its effect is shown in figure~\ref{fig:model_drift_time}.
    \item \textbf{Gaussian~width~\bm{$\sigma$}}:~As described in section~\ref{sec:simulations}, the diffusion of the charge carriers leads to a Gaussian broadening of the charge cloud.
    This can be accounted for in the model by a convolution with a Gaussian distribution with width~$\sigma$.
    The impact of this model parameter on the signal pulse is visualized in figure~\ref{fig:model_gaussian_width}.
    \item \textbf{Exponential~rise~time~\bm{$\tau$}}:~An additional time constant describes the exponential rise of the signal pulses. 
    The parameter factors in the time response of the~CSA and can be interpreted as how quickly charge can be transferred from the detector to the pre-amplifier of the readout circuit~\cite{Spieler2005}.
    The impact of this model parameter is illustrated in figure~\ref{fig:model_exp_rise_time}.
\end{itemize}
\noindent
Combining these four parameters, the signal shape as a function of the time~$t$ can be described by the following model:
\begin{equation}
    \centering
    f(t; A, t_{\mathrm{drift}}, \sigma, \tau) = \left ( \underbrace{A}_{\text{Collected Charge}} \cdot \underbrace{\Theta(t - t_{\mathrm{drift}})}_{\text{Drift Time}} \cdot \underbrace{\left( 1 - \exp(-\frac{t - t_{\mathrm{drift}}}{\tau}) \right)}_{\text{Exponential Rise}} \right ) * \underbrace{\frac{1}{\sqrt{2 \pi} \sigma} \exp(- \frac{t^{2}}{2 \sigma^{2}})}_{\text{Diffusion}}\,.
    \label{eq:signal_model_total}
\end{equation}
This expression is evaluated and fitted to the acquired signal pulses using least squares minimization, see figure~\ref{fig:model_fit}.
The best fit values for~$A$,~$t_{\mathrm{drift}}$,~$\sigma$, and~$\tau$ are extracted for each grid coordinate.
The results are discussed in section~\ref{sec:results}.
Additionally, the \SIrange{10}{90}{\%} signal rise~time~$t_{\mathrm{rise}}$ is also extracted during the fitting procedure.
This enables a comparison of the obtained results with other characterization measurements.
Exemplary, signal pulses with different rise times are depicted in figure~\ref{fig:model_rise_time}.

\FloatBarrier

\section{Results}
\label{sec:results}

\subsection{Determination of the effective charge cloud size}
\label{subsec:cc_width}

\begin{figure}[t]
    \centering
    \captionsetup[subfigure]{justification=centering}
    \begin{subfigure}{0.45\textwidth}
        \centering
        \includegraphics[width=\textwidth,keepaspectratio]{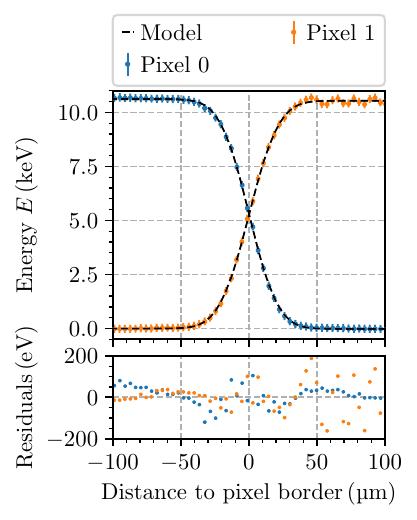}
        \caption[Experiment]{\textbf{Experiment}}
        \label{fig:charge_cloud_width_experiment}
    \end{subfigure}
    \hfill
    \begin{subfigure}{0.45\textwidth}
        \centering
        \includegraphics[width=\textwidth,keepaspectratio]{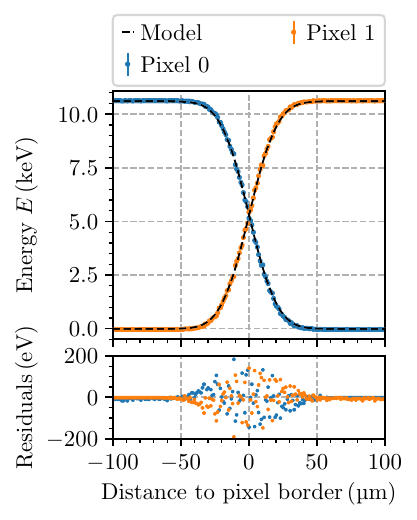}
        \caption[Simulation]{\textbf{Simulation}}
        \label{fig:charge_cloud_width_simulation}
    \end{subfigure}
    \caption[Charge sharing distribution of two adjacent detector pixels]{\textbf{Charge sharing distribution of two adjacent detector pixels}\\
    (a): The charge sharing distribution of the experimental data is shown for pixels~0~(blue) and 1~(orange).
    From the best fit of the model~(black) describing the transition between two detector pixels to the obtained data, the total width of the charge sharing distribution is extracted and afterwards corrected for the width of the laser beam to determine the effective size of the charge cloud.
    The variation of the reconstructed energy in pixel~1 is the result of fluctuations of the laser output.\\
    (b): As in the experiment, the charge sharing distribution in the simulation is obtained between pixels~0~(blue) and~1~(orange).
    The effective size of the charge cloud is extracted from the best fit of the model~(black) to the obtained data.
    The residual structure indicates that the assumed model does not fully describe the charge sharing between two detector pixels.
    }
    \label{fig:charge_cloud_width}
\end{figure}

The effective size of the electron charge cloud is an essential parameter in the model of the detector response to incident radiation.
The parameter is used to quantify the area of a detector pixel that is affected by the effect of charge sharing.
As described in section~\ref{sec:setup}, charge clouds generated at the boundary region between adjacent detector pixels are subject to this effect.
Electrons forming a charge cloud are collected in a certain proportion by each of these pixels.
Therefore, a charge cloud that is generated directly at the boundary is evenly divided between the involved detector pixels.
The sum of the collected charge yields the total deposited energy of the incident radiation since no charges are lost at the boundaries of adjacent pixels.
\newline\indent
The energy deposited in a particular detector pixel is obtained from the amplitude~$A$ of the recorded signals.
Changing the location of the charge cloud formation and the distance to the pixel boundary results in a distribution of shared charge.
Figure~\ref{fig:charge_cloud_width} shows this charge sharing distribution~(CSD) at the border of two adjacent pixels.
The~CSD can be modeled empirically by an integral over a Gaussian distribution with width~$w$ describing the effective size of the charge cloud~(cc).
This model is motivated by the Gaussian spreading of the charge cloud due to diffusion as explained in section~\ref{sec:simulations}.
The figures~\ref{fig:charge_cloud_width_experiment}~and~\ref{fig:charge_cloud_width_simulation} show the~CSDs obtained from the experiment and the simulation, respectively.
Using $\chi^{2}$ minimization, the best fit of the model to both~CSDs yields the effective size of the charge cloud at an energy~$E = \SI{10.6}{\keV}$ for the experiment~(exp) and simulation~(sim):
\begin{align}
    w^{\mathrm{exp}}_{\mathrm{cc}} = \SI{16.3 \pm 0.2}{\micro\meter} \qquad w^{\mathrm{sim}}_{\mathrm{cc}} = \SI{16.3 \pm 0.3}{\micro\meter}\,.
\end{align}
The effective size of the charge cloud~$w^{\mathrm{exp}}_{\mathrm{cc}}$ is calculated from the total width~$w_{\mathrm{tot}}$ of the~CSD obtained from the experimental data.
The total width~$w_{\mathrm{tot}}$ is the quadratic sum of the width of the laser beam and of the charge cloud: $(w_{\mathrm{tot}})^{2} = (w_{\mathrm{laser}})^{2} + (w^{\mathrm{exp}}_{\mathrm{cc}})^{2}$.
Factoring in the beam width of $w_{\mathrm{laser}} = \SI{6.3 \pm 0.7}{\micro\meter}$, we obtain $w^{\mathrm{exp}}_{\mathrm{cc}}$ by averaging the data of about~\si{50}~CSDs.
Both, $w^{\mathrm{exp}}_{\mathrm{cc}}$ and $w^{\mathrm{sim}}_{\mathrm{cc}}$, are well in agreement, which suggests a reliable description of the charge transport in the simulation.
However, the residual structure of both~CSDs shows that the assumed empirical model does not fully describe the charge sharing between two neighboring detector pixels.
A possible explanation could be that neither the effect of self-repulsion nor the long dwell time of the electrons at the saddle point between the pixels is considered in the model.
If only diffusion is activated in the simulation and self-repulsion is deactivated, the effective size of the charge cloud reduces to~$w^{\mathrm{sim, diff. only}}_{\mathrm{cc}} = \SI{14.8 \pm 0.2}{\micro\meter}$.
\begin{figure}[t]
    \centering
    \includegraphics[width=0.95\textwidth,keepaspectratio]{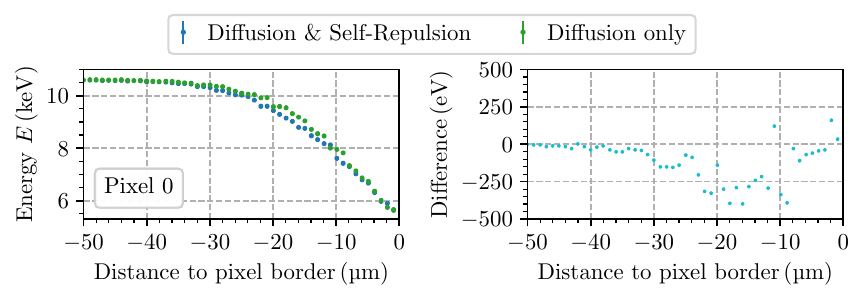}
    \caption[Comparison of simulation scenarios]{\textbf{Comparison of simulation scenarios}\\
    In the simulation, the charge sharing distribution is compared for two different scenarios: (1)~diffusion and self-repulsion being enabled~(blue) and (2)~only diffusion being enabled~(green).
    Compared to scenario~(1), the effective size of the charge cloud is smaller in the diffusion-only case.
    The difference between both scenarios highlights the impact of self-repulsion close to the saddle point in the electric potential.
    }
    \label{fig:charge_cloud_width_simulation_comparison}
\end{figure}
Despite a better description of the simulated data and smaller residuals of best fit of the model to the~CSD~(the reduced $\chi^{2}$ decreases from~$3.2$ for the diffusion and self-repulsion scenario to~$2.4$ for the diffusion only scenario), $w^{\mathrm{sim, diff. only}}_{\mathrm{cc}}$ does neither agree with the experiment nor the simulated data in the presence of both diffusion and self-repulsion.
The largest difference between the two simulation scenarios, is at a distance of~\SIrange{10}{30}{\micro\meter} to the pixel border~(see figure~\ref{fig:charge_cloud_width_simulation_comparison}).
There, i.\,e.\ close to the saddle point in the electric potential, self-repulsion is the most dominant effect of the charge carrier transport.
Therefore, a more comprehensive model needs to be developed in the future that represents the process of charge collection at the pixel boundaries more realistically.
In addition, the charge sharing between three adjacent detector pixels requires more detailed examination, as the saddle point in the electrical potential, and thus the dwelling of electrons, is particularly pronounced here.
Furthermore, it should be investigated whether the effective size of the charge cloud depends on energy and temperature, since the self-repulsion increases quadratically with the number of charge carriers and the mobility increases inversely with the absolute temperature of the detector material.
These investigations will be possible due to the development of an electron gun that provides an electron beam with up to~\SI{25}{\keV} kinetic energy~\cite{Urban2024}.
\FloatBarrier

\subsection{Drift time of electron charge clouds}
\label{subsec:drift_time}

We define the drift time~$t_{\mathrm{drift}}$ of an electron charge cloud as the time difference between the generation of the electron-hole pairs and the arrival of the electrons at the anode of a detector pixel.
The applied bias voltages~(see table~\ref{tab:bias_voltages}) as well as the detector geometry have a significant impact on this property.
Using sidewards depletion in the TRISTAN~SDDs leads to a considerably longer period of electrons drifting in the detector compared to planar detectors or PIN diodes~\cite{Spieler2005}.
Increasing the potential difference between $V_{\mathrm{R1}}$ and $V_{\mathrm{RX}}$ leads to a larger potential gradient between the innermost and outermost drift rings.
This can reduce the drift time of charge clouds by more than~\SI{10}{\nano\second} while still operating the detector within its voltage limits.
At the same time,~$t_{\mathrm{drift}}$ naturally increases with the distance between the interaction position and the pixel anode.

\begin{figure}[t]
    \centering
    \captionsetup[subfigure]{justification=centering}
    \begin{subfigure}{0.49\textwidth}
        \centering
        \includegraphics[width=\textwidth,keepaspectratio]{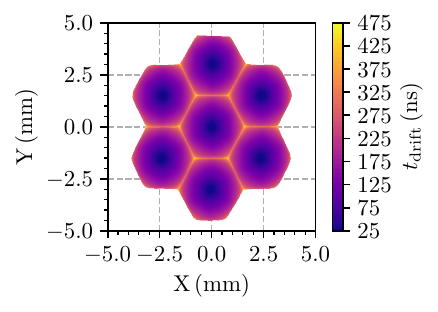}
        \caption[Drift time in all detector pixels]{\textbf{Drift time in all detector pixels}}
        \label{fig:drift_time_seven_pixel}
    \end{subfigure}
    \hfill
    \begin{subfigure}{0.49\textwidth}
        \centering
        \includegraphics[width=\textwidth,keepaspectratio]{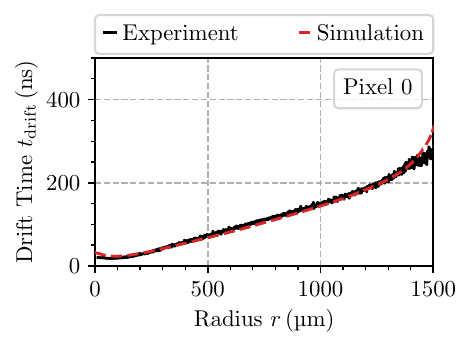}
        \caption[Radial dependence of the drift time]{\textbf{Radial dependence of the drift time}}
        \label{fig:drift_time_central_pixel}
    \end{subfigure}
    \caption[Reconstructed drift times of the electron charge clouds]{\textbf{Reconstructed drift times of the electron charge clouds}\\
    (a): The drift time distribution is shown for all detector pixels.
    The individual pixels can be well recognized as the isochrones resemble the hexagonal structure of the drift rings.\\
    (b): To investigate the radial dependence of the drift time, the two-dimensional drift time distribution is projected to one dimension and averaged for the central detector pixel~(pixel~0).
    In a wide regime, a linear relation between the drift time~$t_{\mathrm{drift}}$ and the interaction radius~$r$ is obtained.
    The simulation agrees to almost full extent to the experimental data.
    }
    \label{fig:drift_time}
\end{figure}

From the scan of the entire detector, the drift time of the created electron charge clouds is determined from the model parameter $t_{\mathrm{drift}}$ of the obtained signals and is shown in figure~\ref{fig:drift_time_seven_pixel}.
Following the pixel arrangement in figure~\ref{fig:tristan_seven_pixel_detector}, the distribution of~$t_{\mathrm{drift}}$ reproduces the individual detector pixels.
The distribution forms isochrones, i.\,e.\ lines representing charge clouds of identical drift time to the respective readout anodes.
Upon closer inspection, it becomes apparent that the isochrones bear a resemblance to the drift ring geometry of the individual pixels.
While charge clouds created near the pixel center are characterized by a drift time of approximately~\SI{25}{\nano\second},~$t_{\mathrm{drift}}$ increases to about~\SI{475}{\nano\second} at the pixel borders, which is compatible to reports by~\cite{Giacomo2022}.

In figure~\ref{fig:drift_time_central_pixel}, the two-dimensional drift time distribution of the central detector pixel~(pixel~0) is projected to one dimension to investigate the radial dependence of the drift time.
Since the drift of an electron charge cloud is mainly determined by its transport in the potential valley~\cite{Lechner1996}, which maintains a linear shape due to the potential gradient, a linear relation is observed between the drift time and the radial position of the interaction of the incident radiation with the detector material in a wide regime.
The average drift time is obtained to compensate for deviations due to the hexagonal shape of the pixels.
In the range of~$r = \SIrange{500}{1000}{\micro\meter}$, a linear model is fitted to the measured data to determine the drift time $t_{\mathrm{drift}}$ as a function of the radius~$r$: 
\begin{equation}
    \centering
    t_{\mathrm{drift}}(r) \propto \SI{149.2 \pm 0.6}{\frac{\nano\second}{\milli\meter}} \cdot r\,.
\end{equation}
The shape differences of the drift time distribution beyond the linear regime can be attributed to several factors.
Charge carriers generated closer to the center pass fewer drift rings, resulting in a reduced impact on the drift time.
Instead, the field components at the pixel center now determine the drift.
At this location, the charge clouds either drift directly to the anode, leading to shorter drift times, or they are guided around the integrated JFET, resulting in slightly increased drift times.
At the largest radial positions, the observed linear behavior also no longer holds.
There, the drift of the charges is affected by the proximity to the pixel boundaries.
The main factor contributing to the noticeable increase in drift time is the extended dwell time of the electrons at the saddle point in the electric potential between adjacent pixels as described in section~\ref{sec:simulations}.

Furthermore, the experimentally obtained radial dependence of the drift time is in good agreement with simulations for almost the entire pixel area.
Despite a slight underestimation in the linear region compared to the mean drift time of the charge clouds, the simulation results reproduce the experimental data to almost full degree.
The deviation at the pixel center can mostly like be attributed to the fact that the integrated JFET is not reflected in simulations realistically.
While the transistor in the simulations is modeled by electrodes on the readout side, in reality the JFET extends a few micrometers into the detector volume.
This variation leads to a modified field distribution and thus to a different description of the charge carrier transport in this region.
In addition, the simulations predict overall higher drift times at the pixel borders compared to the averaged experimental data.
The reason for this is that the simulations are not performed on a grid covering the entire detector pixel as in the experiment, but only in the one-dimensional direction because of computational limitations.
Moreover, the charge clouds collected in this pixel may or may not be affected by charge sharing due to the hexagonal geometry.
Despite the same distance to the pixel center, charge clouds originating from a region close to one of the detector triple points are characterized by shorter drift times than charge clouds affected by charge sharing.
This results in an overall reduced average $t_{\mathrm{drift}}$ at the boundary to adjacent pixels.

\begin{figure}[t]
    \centering
    \begin{subfigure}{0.49\textwidth}
        \centering
        \includegraphics[width=\textwidth,keepaspectratio]{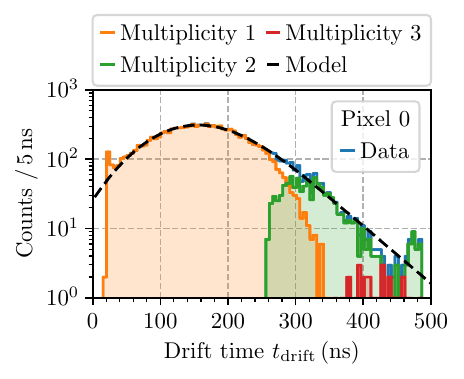}
        \caption[Histogram of the drift time distribution]{\textbf{Histogram of the drift time distribution}}
        \label{fig:drift_time_hist}
    \end{subfigure}
    \hfill
    \begin{subfigure}{0.49\textwidth}
        \centering
        \includegraphics[width=\textwidth,keepaspectratio]{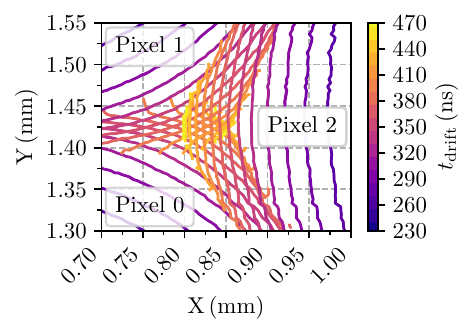}
        \caption[Reconstructed drift time at a triple point region]{\textbf{Reconstructed drift time at a triple point region}}
        \label{fig:drift_time_triple_point}
    \end{subfigure}
    \caption[Relation between the drift time and charge sharing]{\textbf{Relation between the drift time and charge sharing}\\
    (a): For events with a total deposited energy above~\SI{10}{\keV}, the histogram shows the drift times of the electron charge clouds for the central detector pixel~(pixel~0).
    Based on the multiplicity, the tail of the exponentially modified Gaussian distribution can be attributed to charge sharing events with multiplicity 2 and 3.
    In contrast, regular events with multiplicity 1 can be associated to the Gaussian component.
    Events with multiplicity~2 above~\SI{450}{\nano\second} are actually of multiplicity~3, but the energy deposited in the third pixel was below the energy threshold.\\
    (b): Charge clouds of equal drift times are illustrated by isochrones following the drift ring geometry and create concentric lines.
    In the pixel boundary regions, the drift time increases to its maximum values and the intersecting isochrones visualize the area in which charge sharing occurs.}
    \label{fig:drift_time_2}
\end{figure}

The drift time distribution of all events in the central detector pixel with reconstructed energies above \SI{500}{\eV} is shown in the histogram in figure~\ref{fig:drift_time_hist}.
The distinct peak at~$t_{\mathrm{drift}} = \SI{25}{\nano\second}$ corresponds to events with a minimum drift time for all created charge clouds.
Other than that, the histogram follows the shape of an exponentially modified Gaussian distribution~\cite{Grushka1972}.
Fitting this model to the distribution, the sample standard deviation can be extracted.
A value of~\SI{77.0 \pm 2.2}{\nano\second} is obtained as an average across all detector pixels which is an indicator for the intrinsic time resolution of the SDD.
Two consecutive events can produce overlapping signal pulses and thus create pile-up events.
If the difference of the drift times of both charge clouds is less than~\SI{\sim 80}{\nano\second}, the energies of the individual events cannot be determined properly anymore.
Depending on the minimum time resolution of the energy reconstruction timing filter, this will be the limiting factor for precise timing and event reconstruction in the upcoming operation of the TRISTAN detector~\cite{Descher2019}.
A possible mitigation of these limitations could be a different implementation of the event detection and filtering mechanisms, as well as optimization of the electric and magnetic field configurations at the detector section of the~KATRIN beamline.
The latter adjustments are necessary to reduce the number of electrons that are initially backscattered from the detector surface and subsequently backreflected by the electromagnetic fields~\cite{Spreng2024}.

The multiplicity of an event refers to the number of pixels that collected electrons of a charge cloud created in a single event and is therefore an important characteristic of the effect of charge sharing.
The multiplicity reconstruction shows that the Gaussian component of the distribution corresponds to events whose entire charge is collected in a single pixel, hence being attributed to multiplicity~1.
Most of the events have this multiplicity and feature drift times of \SIrange{80}{240}{\nano\second}.
Since the detector surface is uniformly illuminated due to the equidistant grid, this observation is consistent with geometric considerations.
In contrast, the exponential tail of the distribution starting at drift times of approximately~\SI{260}{\nano\second} is due to electrons being collected by two different detector pixels.
Consequently, these events create signals in these two pixels and can thus be classified as multiplicity~2.
These charge clouds are characterized by drift times of about~\SIrange{280}{400}{\nano\second}.
Charge clouds created at a triple point region share their charge carriers between three adjacent pixels and are classified as multiplicity~3 events.
These events feature the highest drift times with values above~\SI{400}{\nano\second}.
Figure~\ref{fig:drift_time_triple_point} shows this region for pixels~0,~1,~and~2.
The triple point itself can be clearly recognized as charge clouds created here are characterized by drift times of up to~\SI{470}{\nano\second}.
In figure~\ref{fig:drift_time_hist}, these can also be identified as the peak at the right end of the distribution.

In summary, knowing the drift time of an electron charge cloud reveals two essential aspects:
First, from the drift time we can deduce the radial position of the generation of the electron-hole pairs in a detector pixel.
Starting with the lowest drift times in the pixel center, and followed by a linear increase almost to the pixel borders, the drift time is an excellent parameter to describe the radial dependence of the charge cloud drift.
Second, the highest drift times at the pixel boundaries allow events affected by charge sharing to be detected and treated separately.
In addition, we can also distinguish whether the charge is shared between two or three neighboring pixels and handle them differently.

\FloatBarrier

\subsection{Event discrimination by signal timing}
\label{subsec:event_discrimination}

The laser measurement allows to assess the precise drift time, as the start of the signal is known via the timed laser pulse.
In regular measurement modes, in contrast, the drift time is not accessible, as the start time of the signal is not known.
In this section, we explore the option of obtaining timing information, and thus the position of the energy deposition via the pulse shape.
To this end, we make use of the additional model parameters~$\sigma$ and~$\tau$, as well as the \SIrange{10}{90}{\%} rise time~$t_{\mathrm{rise}}$.

\begin{figure}[t]
    \centering
    \captionsetup[subfigure]{justification=centering}
    \begin{subfigure}{0.49\textwidth}
        \centering
        \includegraphics[width=\textwidth,keepaspectratio]{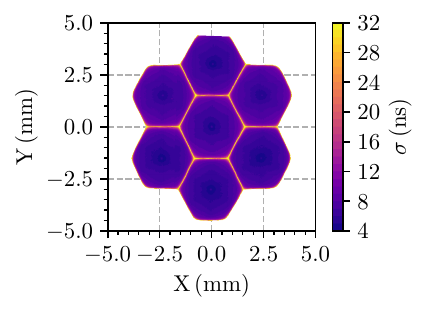}
        \caption[Gaussian width in all detector pixels]{\textbf{Gaussian width in all detector pixels}}
        \label{fig:gaussian_width_seven_pixel}
    \end{subfigure}
    \hfill
    \begin{subfigure}{0.49\textwidth}
        \centering
        \includegraphics[width=\textwidth,keepaspectratio]{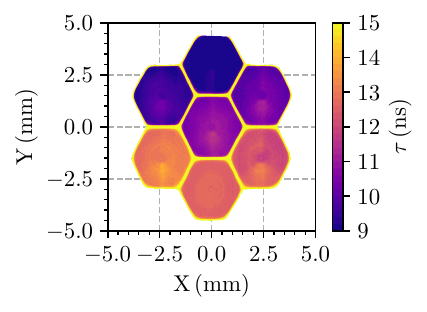}
        \caption[Exponential rise time in all detector pixels]{\textbf{Exponential rise time in all detector pixels}}
        \label{fig:exp_rise_time_seven_pixel}
    \end{subfigure}
    \caption[Distributions of the signal timing parameters]{\textbf{Distributions of the signal timing parameters}\\
    (a): The Gaussian width~$\sigma$ of the signal model follows the hexagonal structure of the drift ring geometry.
    At the pixel borders, this behavior is altered due to the presence of a saddle point in the electric potential between adjacent pixels.\\
    (b): The exponential rise time~$\tau$ is approximately constant for the individual pixels.
    The gradient of~$\tau$ from the upper to the lower half of the detector can be explained by an additional capacitive coupling for the longer readout lines.
    The highest charge collection times in the SDD and therefore the highest values of~$\tau$ are the result of electrons having an extended dwell time at the boundaries between adjacent pixels.
    }
    \label{fig:timing_constants}
\end{figure}

Figures~\ref{fig:gaussian_width_seven_pixel}~and~\ref{fig:exp_rise_time_seven_pixel} show the spatial, two-dimensional distributions of both model parameters~$\sigma$ and~$\tau$.
As the Gaussian spreading of the charge cloud during the drift is proportional to~$\sqrt{t_{\mathrm{drift}}}$, the distribution of~$\sigma$ is comparable to the one of the drift time, cf.\,~figure~\ref{fig:drift_time_seven_pixel}.
The reconstruction clearly shows the hexagonal ring structure and is homogeneous for all detector pixels.
In contrast, the exponential rise time~$\tau$ of the signal pulses is independent of~$t_{\mathrm{drift}}$.
Instead, it is nearly constant within one pixel, but changes depending on the position of the pixel on the silicon wafer.
While the top pixel has values of~$\tau \leq \SI{9}{\nano\second}$, the bottom pixel has~$\tau \geq \SI{13}{\nano\second}$.
Moreover, the pixels in between steadily increase by about~\SI{1}{\nano\second} from top to bottom.
This behavior can be explained by variations of the capacitive load on the signal and biasing lines of the metal bus structure on the readout side of the detector.
As shown in figure~\ref{fig:tristan_seven_pixel_detector}, these lines run from the top to the bottom.
Consequently, pixels on the lower half have longer signal lines and therefore an increased capacitive coupling resulting in higher exponential rise times.
Since this effect also depends on the dimensions of the bus structure, the width of the metal strips has been reduced compared to previous iterations of the detector design.
In addition, the spacing between the lines is increased to minimize their coupling and thus reduce crosstalk.

In the region of the pixel borders where the charge sharing occurs, the Gaussian width parameter increases to values of~$\sigma > \SI{10}{\nano\second}$ due to the long dwell time at the saddle point.
Over time, parts of the charge cloud drift towards the potential minimum at the anode, while other parts remain at the saddle point and follow only gradually.
In contrast to diffusion which leads to an isotropic expansion of the charge cloud, this causes the cloud to stretch in the radial direction.
As a result, the first electrons arrive at the anode only after a drift time of about~\SI{300}{\nano\second}.
At the same time, this behavior leads to an extended charge collection time, which is reflected in a prolonged exponential rise of the signal pulses with values of~$\tau \geq \SI{15}{\nano\second}$.

Both effects can also be observed in the reconstructed \SIrange{10}{90}{\%}~signal rise time.
\begin{figure}[t]
    \centering
    \captionsetup[subfigure]{width=0.9\linewidth, format=hang}
    \begin{subfigure}{0.49\textwidth}
        \centering
        \includegraphics[width=\textwidth,keepaspectratio]{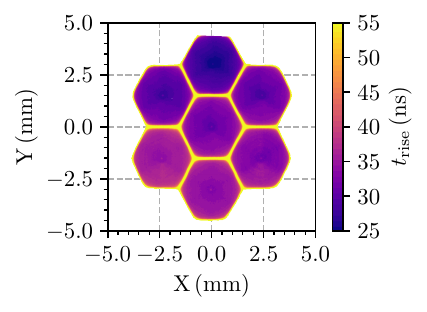}
        \caption[\SIrange{10}{90}{\%} signal rise time in all detector pixels]{\textbf{\SIrange{10}{90}{\%} signal rise time in all detector pixels}}
        \label{fig:rise_time_seven_pixel}
    \end{subfigure}
    \hfill
    \begin{subfigure}{0.49\textwidth}
        \centering
        \includegraphics[width=\textwidth,keepaspectratio]{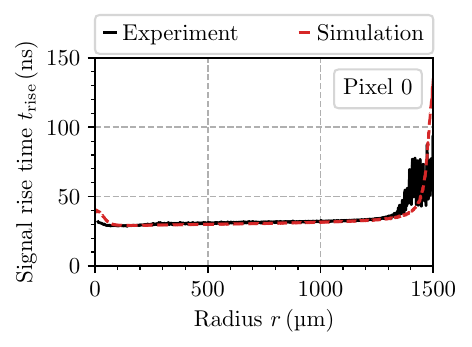}
        \caption[Radial dependence of the \SIrange{10}{90}{\%} signal rise time]{\textbf{Radial dependence of the \SIrange{10}{90}{\%} signal rise time}}
        \label{fig:rise_time_1d}
    \end{subfigure}
 \caption[Distributions of the signal timing parameters]{\textbf{Distributions of the signal timing parameters}\\
    (a): As a combination of the model parameters~$\sigma$~and~$\tau$, the \SIrange{10}{90}{\%}~signal rise time~$t_{\mathrm{rise}}$ enables the identification of charge sharing events at pixel borders and thus the discrimination against regular events.\\
    (b): For the central detector pixel~(pixel 0), the experimental data is averaged and the radial dependence of~$t_{\mathrm{rise}}$ is shown.
    A linear relation between the interaction radius~$r$ and $t_{\mathrm{rise}}$ is observed in a wide range.
    The simulation is in high agreement with the experimental data.
    }
    \label{fig:signal_rise_time}
\end{figure}
Figure~\ref{fig:rise_time_seven_pixel} shows the corresponding spatial distribution of~$t_{\mathrm{rise}}$.
We find that the \SIrange{10}{90}{\%} signal rise time is a combination of the two timing model parameters.
This can be related to the fact that the hexagonal-shaped distribution of~$\sigma$ as well as the overall gradient following~$\tau$ can be recognized in the reconstruction of~$t_{\mathrm{rise}}$.
Consequently, the signal rise time can also be used as a parameter to describe the radial position of the interaction location in the detector as illustrated in figure~\ref{fig:rise_time_1d}.
In the same range of~$r = \SIrange{500}{1000}{\micro\meter}$ used for the drift time in section~\ref{subsec:drift_time}, a linear relation is determined for the \SIrange{10}{90}{\%} signal rise time~$t_{\mathrm{rise}}$ as function of the radius~$r$:
\begin{equation}
    \centering
    t_{\mathrm{rise}}(r) = \SI{2.4 \pm 0.1}{\frac{\nano\second}{\milli\meter}} \cdot r + c\,.
\end{equation}
As the constant~$c$ depends on the length of the readout lines of the metal bus structure and therefore on~$\tau$, higher rise time values are observed towards the lower half of the detector.
For the central pixel,~$c$ is determined to~\SI{30.3 \pm 0.1}{\nano\second} which is compatible with the determined range of~$\SI{25}{\nano\second} < c < \SI{35}{\nano\second}$.
Although the simulation is highly consistent with the averaged experimental data in the linear region, the deviations at small radii can be attributed to the drift of charge clouds near the integrated~JFET in the center of the pixel as described in section~\ref{subsec:drift_time}.
In contrast, compared to the drift time at large radii, $t_{\mathrm{rise}}$ is higher than predicted in the simulations.
This phenomenon can be attributed to the averaging of charge sharing events with longer charge collection times and regular events with shorter collection times, despite the fact that the distance to the pixel center is the same.

\begin{figure}[t]
    \centering
    \captionsetup[subfigure]{width=0.9\linewidth, format=hang}
    \begin{subfigure}{0.49\textwidth}
        \centering
        \includegraphics[width=\textwidth,keepaspectratio]{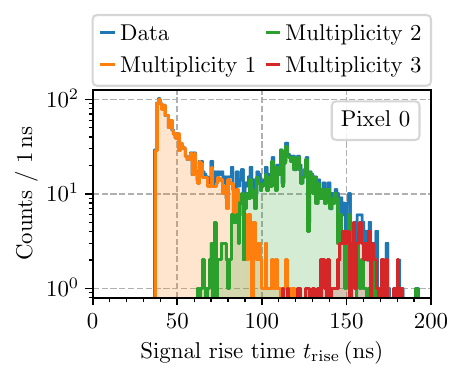}
        \caption[Histogram of the rise time distribution at a detector triple point]{\textbf{Histogram of the rise time distribution at a detector triple point}}
        \label{fig:rise_time_histogram}
    \end{subfigure}
    \begin{subfigure}{0.49\textwidth}
        \centering
        \includegraphics[width=\textwidth,keepaspectratio]{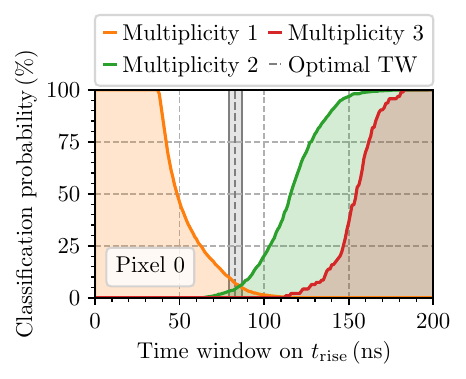}
        \caption[Event classification based on a signal rise time window]{\textbf{Event classification based on a signal rise time window}}
        \label{fig:rise_time_cut}
    \end{subfigure}
    \caption[\SIrange{10}{90}{\%}~signal rise time and the identification of charge sharing]{\textbf{\SIrange[detect-weight]{10}{90}{\%}~signal rise time and the identification of charge sharing}\\
    (a): In the same region as shown in figure~\ref{fig:drift_time_triple_point}, the rise time distribution for events with energies above \SI{500}{\eV} is reconstructed and the histogram is determined for the central detector pixel~(pixel~0).
    Regular events with multiplicity~1 are characterized by the lowest signal rise times up to about~\SI{100}{\nano\second}.
    The sharp drop at the lower end is the result of the limited area examined during the laser scan.
    While charge sharing events with multiplicity~2 are dominant at rise times between~\SI{90}{\nano\second} and~\SI{140}{\nano\second}, those with multiplicity~3 are prevalent between~\SI{140}{\nano\second} and~\SI{180}{\nano\second}.\\
    (b): To distinguish between regular and charge sharing events, a time window could be defined which classifies events based on their~$t_{\mathrm{rise}}$.
    The vertical dashed line shows the optimal time window for which almost all charge sharing events with multiplicity~2~and~3 could be rejected, while nearly all regular events with multiplicity~1 would be accepted.
    }
    \label{fig:rise_time_event_identification}
\end{figure}

Within each pixel, the \SIrange{10}{90}{\%} rise time of the signal pulses is spread only over about~\SI{5}{\nano\second}.
However, at the pixels borders signal rise times well above~\SI{50}{\nano\second} are measured.
For the central detector pixel, the rise time distribution at the scanned detector triple point~(compare to figure~\ref{fig:triple_point_grid}) is shown in the histogram in figure~\ref{fig:rise_time_histogram}.
Charge sharing events are observed starting at signal rise times of about~\SIrange{60}{70}{\nano\second} and increasing almost up to approximately~\SI{200}{\nano\second} at the boundary between adjacent pixels.
This distinct increase results from the prolonged exponential rise of the signal pulses.
Events with multiplicity~2 are usually characterized by signal rise times between~\SI{90}{\nano\second} and~\SI{140}{\nano\second}.
At the detector triple point, the signal rise time reaches typical values between~\SI{140}{\nano\second} and~\SI{180}{\nano\second}.
Consequently, the signal rise time can be used to identify the nature of an event.
Exemplary, this method is illustrated in figure~\ref{fig:rise_time_cut} for the central detector pixel.
The objective is to define a time window on the signal rise time that minimizes the probability of misclassifying a charge sharing event as a regular event and vice versa.
As the time window increases, the probability of correctly categorizing events with multiplicity~1 as regular events decreases. 
In contrast, the probability increases for events with multiplicity~2~and~3 being correctly identified as charge sharing events.
The optimum time window is determined to~\SI{83 \pm 4}{\nano\second} as an average across the scanned pixels~0,~1,~and~2.
Applying this threshold on the signal rise time of the detected events allows to reject~\SI{97 \pm 2}{\%} of the charge sharing events.
At the same time, \SI{95 \pm 2}{\%} of the regular events would be accepted for further processing such as energy evaluation.
The simulations predict that~$t_{\mathrm{rise}}$ is almost independent of the deposited energy in the investigated range of~\SIrange{10}{60}{\keV}.
However, these results should be verified experimentally, as the effect of the detector entrance window, which mainly affects electrons, is not taken into account in the simulations~\cite{Siegmann2024}.
Nevertheless, for the upcoming operation of the TRISTAN detector, this procedure makes~$t_{\mathrm{rise}}$ a potential parameter for the event selection as well as for the discrimination of charge sharing events with multiplicity~2 and~3.
\FloatBarrier

\section{Conclusions and outlook}

In this work, we successfully characterized a 7-pixel~TRISTAN prototype silicon drift detector with a laser system in a dedicated experimental setup.
The effect of charge sharing as well as timing characteristics, especially the drift time and the~\SIrange{10}{90}{\%}~signal rise time~$t_{\mathrm{rise}}$, were studied in detail.
For different interaction positions in the detector, the signal pulses were analyzed using an empirical model and relevant parameters were extracted.
Furthermore, the experimental results were validated with simulations and found to be in high agreement.

First, the effect of charge sharing at the boundaries of adjacent detector pixels was investigated.
The effective size of the charge cloud, which is an essential parameter for the final description of the detector response model, was determined from the study of the transition region between two detector pixels.
However, it was found that the empirical model describing this effect was not detailed enough to account for the processes such as self-repulsion or the long dwell time of electrons at the saddle point of the electric potential.
Therefore, it is necessary to revisit this empirical model.

The reconstruction of the drift time of the electron charge clouds has shown that the hexagonal drift ring geometry of the detector pixels determines the drift of the electrons.
The drift time was found to be an excellent parameter to describe the radial dependence of the charge cloud drift.
It was also possible to determine the intrinsic time resolution of the~SDD, which together with the time resolution of the energy filter is the critical constraint for temporal event separation.
The homogeneity of the signals in all detector pixels has been investigated and small differences could be explained by the detector layout and signal routing.

Moreover, the feasibility of event discrimination based on precise signal timing was demonstrated.
With the definition of a time window acting as a threshold, $t_{\mathrm{rise}}$ can be used for the classification of regular and charge sharing events.
Only events with signal rise times below this threshold would be accepted in the event selection process and others can be rejected from the analysis.
This makes the~\SIrange{10}{90}{\%}~signal rise time a reliable indicator for the identification of charge sharing events at the pixel borders.

In order to facilitate the operation within the~KATRIN beamline, the 7-pixel prototype detectors are being scaled up to 166-pixel detector modules.
It is anticipated that the planned~SDD focal plane array will eventually consist of a 9-module system.
As a consequence of the increased dimensions of the~SDD chips, it is necessary to validate the experimental outcomes and to direct particular attention to the signal propagation times.
In consideration of the extended signal readout components and the overall increased length of cables, it is expected that higher signal rise times will be observed, which will result in a shift of the time window for the detection of regular events and charge sharing events.

In summary, probing the surface of a 7-pixel prototype~TRISTAN~SDD with a laser system allowed the detailed study of the charge carrier drift in the detector pixels and at their boundaries.
The investigations gave valuable insights on the signal timing and its relation to the detector geometry.
The experimental results act as input parameters for sensitivity studies and are an important contribution to the model of detector response for the upcoming~\si{\keV}-scale sterile neutrino search with~KATRIN.

\acknowledgments

We acknowledge the support of Ministry for Education and Research BMBF~(05A20PX3), Max Planck Research Group~(MaxPlanck@TUM), Deutsche Forschungsgemeinschaft DFG Graduate School grant no.~SFB-1258, Excellence Cluster ORIGINS in Germany, and Istituto Nazionale di Fisica Nucleare~(INFN, Italy)~CSN2.
This project has received funding from the European Research Council~(ERC) under the European Union Horizon 2020 research and innovation programme (grant agreement no.~852845).
Our special thanks are also due to S.~Schmidl of the Max Planck Institute for Physics for his support and assistance in the commissioning of the laser system.

\bibliographystyle{JHEP}
\bibliography{biblio.bib}

\end{document}